\documentstyle{article}
 \date{}                                                   

 \newcommand{\beq}{\begin{equation}}                       
 \newcommand{\eeq}{\end{equation}}                         
 \newcounter{nt}[section]                                  
 \newcounter{nl}[section]                                  
\begin{document}

 \title{ {\bf Asymptotic analysis for the strip problem related to a
 parabolic third - order operator}}

 \author{{\sc M. De Angelis} \\ {\small Dep of Math and
 Appl, University of Naples,} \\{\small via Claudio 21, 80125, Naples, Italy.}}

 \maketitle

\vspace{3mm}

\begin{center}
 \parbox{13cm}{{\footnotesize{\normalsize\bf
 Abstract---}}\footnotesize Aim of this paper
 is the qualitative analysis of a boundary value problem for a third
 order non linear parabolic equation
 which describes several dissipative models. When the source term
 is linear, the problem is explictly solved by means of a Fourier series
 with properties of rapid convergence. In the non linear case,
 appropriate estimates of this series allow to deduce
 the asymptotic behaviour of the solution.\vspace{3mm}
}
\end{center}

 \begin {center}
 \section{\hspace*{-6mm}{\large\bf .\hspace{2mm}INTRODUCTION}}
 \end {center}
 \setcounter{equation}{0}

The evolution of several dissipative phenomena can be described by the non linear equation

 \beq                                   \label{11}
 {\cal L}_ \varepsilon u= (\partial_{xx}(\varepsilon
\partial_t+c^2 ) - \partial_t(\partial_t+a))u=
F(x,t,u),
 \eeq

\vspace{1mm}\noindent
where  $ \varepsilon, a, c$ are positive constants and
 $F=F(x,t,u)$  is a prefixed function.
For instance, the equation (\ref{11}) governs
 sound propagation in
viscous gases as reported in the classical literature [\ref{lam}], motions of viscoelastic fluids or solids
[\ref{jrs}], or motions of heat condution at low temperature (see,
e.g., [\ref{mps}] and refecence
therein).
Moreover,
in
superconductivity, (\ref{11}) models the flux dynamics in the
Josephson junction. In this case, let $u=u(x,t)$ be the phase difference of the wave functions related to the two
  superconductors, and let $\gamma$ be the normalized current bias; when
  $F=\sin u -\gamma$, equation (\ref{11}) is the perturbed Sine -
  Gordon equation (PSGE). Terms $\varepsilon u_{xxt}$ and $a u_t$ characterize the
 dissipative normal electron current flow along and across the junction
  and represent the
{\em perturbations} with respect to the classic Sine Gordon
 equation [\ref{sco}]. Moreover, the value's range for a and
 $\varepsilon$ depends on the real junction. In fact, there are
 cases with $0<a,\varepsilon<1 $ [\ref{pag1}] and,
 when the resistence of the junction is so low to
 short completely the
 capacitance, the case a large with respect
 to 1 arises [\ref{tin}].
 As it is well known, up to today, Josephson junctions have
found a lot of applications in many fields and lately
new
high-$T_C $ materials imply an increasing evolution of the
superconducting technologies [\ref{cp}].

If $\ell$ is
the normalized length of the junction, boundary conditions
for
$x=0$ and $x=\ell$  specify the applied external magnetic field
 [\ref{for}, \ref{lom1}]. Then, if

\vspace{3mm}
\begin{center}
$\Omega =\{(x,t) : 0  < x  <
\ell, \  \ 0 < t \leq T \}$,
\end{center}

\noindent
the strip problem for (\ref{11}) can be stated as follows:

  \beq                                                     \label{12}
  \left \{
   \begin{array}{ll}
    & \partial_{xx}(\varepsilon
u _{t}+c^2 u) - \partial_t(u_{t}+au)=F(x,t,u),\ \  \
       (x,t)\in \Omega,\vspace{2mm}\\
   & u(x,0)=g_0(x), \  \    u_t(x,0)=g_1(x), \  \ x\in [0,\ell],\vspace{2mm}  \\
    & u(0,t)=0, \  \ u(\ell,t)=0, \  \ 0<t \leq T.
   \end{array}
  \right.
 \eeq

\vspace{3mm}
 The linear case($F=f(x,t)$) has already solved in
 [\ref{ddr}, {\ref{bari}]
by means of convolutions of
Bessel
functions. However, it seems difficult to deduce an exhaustive
asymptotic analysis by means of this solution.

In this paper, both linear and non linear problems are analyzed and the
Green function of the linear case is determined by Fourier series:

\beq                                                \label{13}
G(x,t,\xi)=\frac{2}{l}\sum_{n=1}^{\infty}
H_n(t) \  \  \sin\gamma_n\xi\  \ \sin\gamma_nx,
\eeq

\noindent
where

\beq                                             \label{14}
\ \  H_n(t)= \frac{1}{\omega_n}e^{-h_nt}
\sinh(\omega_nt), \\ \  \ \gamma_n=\frac{n\pi}{l}, \eeq
\hspace{4cm}\[ b_n=c\gamma_n,\ \ \  h_n=\frac{1}{2}(a+\varepsilon\gamma_n^2),\
 \ \ \omega_n=\sqrt{h_n^2-b_n^2}.\]

This series is characterized by properties of rapid convergence and
allows to establish an exponential decrease of the solution in the linear
case. Moreover, the non linear problem is reduced to
an integral equation with kernel G and, by means of suitable
properties of G, the asymptotic analysis of
the solutions is achieved.

 \begin {center}
 \section{\hspace*{-6mm}{\large\bf .\hspace{2mm} PROPERTIES OF THE
 GREEN FUNCTION}}
 \end {center}

 \setcounter{equation}{0}

Firstly, let us prove that function G defined by
 series (\ref{13})
decreases at least with  $n^{-2}$ and that it
is exponentially vanishing as $t\rightarrow\infty$. In fact, if we put:

\beq
p=
\frac{c^2}{\varepsilon+a(\ell/\pi)^2},  \ \ \ \ \ q=\frac{a+\varepsilon
(\pi/\ell)^2}{2}, \ \ \ \ \  \beta\equiv \min (p,q),
\eeq
the following lemma holds:

\vspace{3mm}
{\bf Lemma 2.1}-
{\em The function} $G(x,\xi,t)$ {\em defined in}
(\ref{13}) {\em and all its time derivatives are continuous functions in}
$\Omega$, $\ \forall \ a, \varepsilon, c^2 $ $\in$  $\Re^+$.
{\em  Moreover, everywhere in } $\Omega$, {\em  it results:}
\beq                                        \label{21}
|G(x,\xi,t)|\,\leq \,M  \,e^{-\beta t}, \ \ \, \ \ \ \
|\frac{\partial ^j G}{\partial t^j}|\, \leq \,N_j \,  e^{-\beta
t},
\\ \ \ \ \ \ \, \ \ j \in {\sf N}
\eeq

\noindent
{\em where } $M, N_j $ { \em are constants depending on } $a,\varepsilon,
c^2$.

{\bf Proof}. When $c^2=a\varepsilon$, the operator $L_\varepsilon$ can be
reduced to the wave operator. If $c^2< a\varepsilon$, for all
$n\, \geq \,1$,
one has $h_n > b_n$.
So, if $k$ is an arbitrary constant such that
$c^2/a\varepsilon\, <\,k\,<1$, Cauchy inequality assures
that:

\beq                                         \label{22}
\frac{b_n}{\sqrt{k}} \ \leq \frac{1}{2} \ (\varepsilon\gamma_n^2+a)=h_n.
\eeq
\noindent
Then, if $X_n =(b_n/h_n)^2$, it results:

\beq
(1-k)^{\frac{1}{2}}\, \leq\,
\frac{\omega_n}{h_n}=(1-X_n)^{\frac{1}{2}}<1-X_n/2.
\eeq
\noindent
So, considering that:

\beq                                       \label{23}
\frac{b_n^2}{2h_n}\, \\ \geq \
p,\ \ \ \ \ \ \ \
 h_n> n^2 \,(q-a/2),  \ \ \ \ \forall
 n \ \geq 1,
\eeq

\noindent
$e^{-t(h_n-\omega_n)}\leq  e^{-pt}$ and $  \omega_n \leq
\frac{(1-k)^{-\frac{1}{2}}}{q-a/2}\, \frac{1}{n^2}$ hold.
As consequence,
 $(\ref{21})_1$ follows.

As for
$(\ref{21})_2$, we observe that  $ h_n-\omega_n \leq \frac{2c^2}{\varepsilon},\
\forall n\geq 1$,
and by means of standard computations  $(\ref{21})_2$ is deduced.

It may be similarly proved that estimates (\ref{21}) hold also when
$c^2>a\varepsilon$. \hbox{} \hfill \rule{1.85mm}{2.82mm}

As for the x-differentiation of Fourier's series like
(\ref{13}), attention must be given to convergence problems. Therefore, we
 consider x-derivatives of the operator
$(\varepsilon\partial_t+c^2)G$       instead of   $G $ and   $G_t$.

\vspace{3mm}
{\bf Lemma 2.2}.- {\em For all} $a ,
\varepsilon, c^2 $ $\in$ $\Re^+$,
{\em the function}   $ G(x,\xi,t)$ {\em defined in} (\ref{13}),
 {\em is such that:}

\beq                            \label{24}
|\varepsilon\, G_t+c^2G| \\ \, \leq A_o \, \,  e^{-\beta t}
,\ \ \ \ |\partial_{xx}(\varepsilon\, G_t\, +c^2\, G )|\\ \,\leq A_1 \, \,
\, e^{-\beta t},
\eeq

\noindent
{\em where $A_i$ $(i=0,1)$ are constants depending  on $a, \varepsilon ,
c^2 $.}

{\bf Proof}.
As for the
hyperbolic terms in $H_n, $ it results:
\beq                                    \label{25}
\ \ \ \ \ \ \ \ \ \varepsilon
\dot H_n+c^2H_n=\frac{1}{2\omega_n}e^{-h_nt}\{(c^2+\varepsilon\varphi_n)
e^{\omega_nt}-[c^2-\varepsilon(h_n+\omega_n)]e^{-\omega_nt}\},
\eeq
\noindent
with
$\varphi_n=h_n(-1+\sqrt{1-X_n})$ and $X_n=(\frac{b_n}{h_n})<1$.

\vspace{1mm}
\noindent
Further, by means of Taylor's formula it's possible to prove  that:

\beq                         \label{26}
|c^2+\varepsilon \ \varphi_n | \ \, \leq \frac{1}{n^2} \
(5\alpha^2/4+\alpha_1/8),
\ \ \ \ (\alpha, \alpha_1 \ = \hbox{const}).
\eeq

\noindent
Estimates of Lemma 2.1 together with (\ref{26}) show that the
terms of the series related to the operator $ (\ref{24})_1$ decrease at
least as $n^{-4}$. So, it can be differentiated term by terms with respect
to x and estimate$ (\ref{24})_2$ holds.
\hbox{}\hfill\rule{1.85mm}{2.82mm}

We mean as solution of the equation
${\cal L}_\varepsilon v
 = 0$  a continuous function
$v(x,t)$ which has continuous the derivatives $v_t, v_{tt}, $ $\partial_{xx}(\varepsilon
v_t+c^2v) $  and these derivatives verify the equation. Therefore,
 the following theorem holds:

\vspace {3mm}
{\bf Theorem 2.1}- {\em The
function  $G=G(x,t)$ defined in} (\ref{13}){\em is a
solution of the equation }

\beq                      \label{27}
{\cal L}_\varepsilon G \, =\partial_{xx}(\varepsilon
G _{t}+c^2 G) - \partial_t(G_{t}+aG)=0.
\eeq
\hbox{}\hfill\rule{1.85mm}{2.82mm}

To obtain the explicit solution of (\ref{12}),
 properties for the convolution of function G with data have to be
analized.

\noindent
For this, let $g(x) $ be a continuous function on $(0,\ell)$ and consider:

\beq                             \label{28}
u_g(x,t)=
\int_{0}^{\ell} g(\xi) \ \ G(x,\xi,t)\ \ d\xi,\ \ \ \ u_g^*(x,t) =
(\partial_t+a-\varepsilon\partial_{xx})u_g(x,t).
\eeq

\noindent
Then one has:

\vspace{3mm}
{\bf Lemma 2.3}- {\em If} $g(x)$ {\em is a} $C^1[0,\ell],$
{\em then
the function} $u_g$  {\em is a solution}
{\em  of the
equation} ${\cal L}_\varepsilon =0$ {\em such that}:

\beq                            \label{29}
\lim_{t \rightarrow 0}u_g(x,t)=0, \ \ \ \ \ \lim_{t \rightarrow 0} \partial_t
u_g(x,t)=g(x),
\eeq

\noindent
{\em uniformly for all }$x\in [0,\ell]$.

{\bf Proof.} Lemma (\ref{21})-(\ref{22}) and continuity of g assure
that function
$(\ref{28})_1$ and partial derivatives required by the solution converge absolutely for all
$(x,t)\in \Omega$. Hence, since Theorem 2.1,  ${\cal L}_\varepsilon u_g =0$.
Besides, hypotheses on the function
g and (\ref{21}), imply $(\ref{29})_1$.
Further, one has:
\beq                            \label{210}
G_t=-\frac{2}{\pi} \ \ \frac{\partial}{\partial\xi} \ \ \sum_{n=1}^{\infty}
\ \ \dot H_n(t) \frac{\cos \gamma_n\xi}{n}\sin
\gamma_n x,
\eeq

\noindent
and hence:

\beq                            \label{211}
\partial_tu_g= -\frac{2}{\pi}
\sum_{n=1}^{\infty}
\ \ \frac{\dot H_n(t)}{n} \ \ [g(\xi)\cos \gamma_n\xi \
]^{\ell}_0 \ \ \sin\gamma_n x \  \eeq
\hspace{4cm}\[+\frac{2}{\pi} \int_{0}^{\ell}\sum_{n=1}^{\infty}
\ \ \dot H_n(t)  \,g^{'}(\xi) \, \frac{\cos \gamma_n\xi}{n} \,\sin
\gamma_n x \ d\xi.\]
So, if $\eta(x)$ is the Heaviside function, from (\ref{211}), one obtains:

\beq                            \label{212}
\ \ \ \ \ \ \ \ \ \lim_{t \rightarrow 0}\partial_tu_g=
\frac{x}{\ell}[g(\ell)-g(0)]+g(0)- \int_{0}^{\ell} g^{'}(\xi)
[\eta(\xi-x)+\frac{x}{\ell}-1]
d\xi  =g(x).
\eeq
\hbox{}\hfill\rule{1.85mm}{2.82mm}

{\bf Lemma 2.4}- {\em If} $g(x) \in C^3[0,\ell]$ {\em with}
$g^{(i)}(0)=g^{(i)}(\ell)=0
\ \  (i=1,2,3)$,
{\em
then the function} $u_g^*$ {\em defined in }$(\ref{28})_2$ {\em is a solution
 of the
equation} ${\cal L}_\varepsilon =0$ {\em such that}:
\beq                            \label{213}
\lim_{t \rightarrow 0}u_g^*(x,t)=g(x), \ \ \ \ \ \lim_{t \rightarrow 0} \partial_t
u_g^*(x,t)=0,
\eeq

\noindent
{\em uniformly for all }$x\in [0,\ell]$.

{\bf Proof.} Hypoteses on $g(x)$
 assure that:

\beq                               \label{214}
\partial_{xx}u_g(x,t) =
\int_{0}^{\ell} g^{''}(\xi) \ \ G(x,\xi,t)\ \ d\xi =  u_{g^{''}}(x,t).
\eeq

\noindent
So, by Lemma 2.3 and expression (\ref{214}), equation  ${\cal
L}_\varepsilon u_g^* =0$ is verified.
Moreover,
since $ \partial_t u_g^* =
c^2u_{g^{''}}$ and
$(\ref{29})_1$, $(\ref{213})_2 $ holds.
Finally, owing to (\ref{29})-(\ref{214}), one has  $(\ref{213})_1$,
too.
\hbox{}\hfill\rule{1.85mm}{2.82mm}

 \vspace{3mm}

 \section{\hspace*{-6mm}{\large\bf .\hspace{2mm}EXPLICIT SOLUTION OF
 THE LINEAR PROBLEM}}

 \setcounter{equation}{0}

Consider the linear problem.
When $F \equiv 0 $ , by Lemma 2.3 and 2.4, one has:

\vspace{3mm}
{\bf Theorem 3.1}- {\em If the initial data}  $g_1(x),$ {\em and }$   g_0(x)$ {\em verify
 hypotheses of Lemma} 2.3 and 2.4, {\em  respectively,} {\em  then
 the homogeneous problem} (\ref{12}) {\em admittes the following solution}:

\beq                        \label{31}
u_0(x,t)=u_{g_1}+(\partial_t+a-\varepsilon\partial_{xx})u_{g_0}.\ \
\eeq
\hspace{14.5cm}\rule{1.85mm}{2.82mm}

As for the linear non-homogeneous problem, ($F \equiv f(x,t) $), let:

\beq                             \label{32}
u_f(x,t)\ \ = - \int_{0}^{t} d\tau
\int_{0}^{\ell} f(\xi,\tau) \ \ G(x,\xi,t-\tau)\ \ d\xi.
\eeq

By means of standard computations, one can verify that the function
(\ref{32}) satisfies (\ref{12}) with $g_0=g_1=0$. As consequence, one
has:

\vspace{3mm}
{\bf Theorem 3.2}- {\em If }$f$ {\em and} $f_x$ {\em
are continuous functions
in $\Omega$,}
 {\em then the function} $u= \ u_0 +u_f$
{\em represents a solution} {\em of the linear non-homogeneous strip
problem.}

{\bf Proof.} Considering that:
\beq
\lim_{\tau \rightarrow t}
 \int_{0}^{\ell} f(\xi,\tau) G_t(x,\xi,t-\tau) d\xi = f(x,t),
 \eeq
Theorem 2.1 assure that ${\cal
L}_\varepsilon u_f =f(x,t)$.

Moreover, since:

\beq                              \label{33}
|u_f| \leq   B_1 (1-e^{-\beta t}); \ \ \ \\ \  |\partial_tu_f| \leq
B_2 (1-e^{-\beta t}), \ \ \ (B_1,B_2 \ = \hbox{const.}),
\eeq

\noindent
 initial homogeneous conditions are satisfy.
\hbox{}\hfill\rule{1.85mm}{2.82mm}

Uniqueness is an obviously consequence of the energy-method (see, for
example, [\ref{dr}]). So, the following theorem holds:
:

\vspace{3mm}
{\bf Theorem 3.3}- {\em When the source term }  $f(x,t)$ {\em satisfies
Theorem} 3.2, {\em and the initial data} $(g_0,g_1)$ {\em satisfy
Theorem} 3.1,
 {\em then the function}

\beq                                \label{34}
u(x,t)=u_{g_1}+(\partial_t+a-\varepsilon\partial_{xx})u_{g_0}+u_f
\eeq
{\em is the unique solution} {\em of the linear non-homogeneous
strip problem} {\em with} $F=f(x,t)$.
\hbox{} \hfill \rule{1.85mm}{2.82mm}

\vspace{1mm}
Consider, now, the {\em non-linear} problem (\ref{12}), and observe
that (\ref{34}) implies:

\beq                                          \label{35}
\ \ \ \ \  \\\\\ \ \ u(x,t)=\,
\int_{0}^{\ell} g_1(\xi) G(x,\xi,t) d\xi
+
(\partial_t+a-\varepsilon\partial_{xx})\int_{0}^{\ell} g_0(\xi)
G(x,\xi,t) d\xi
\eeq
\hspace*{2cm}\[- \int_0^ t d\tau\, \int_0^\ell\, G(x,\xi,t-\tau)\,
F(\xi,\tau,u(\xi,\tau))d\xi.\]

\noindent
which represents an integral equation for u(x,t).

 \begin{center}
 \section{\hspace*{-6mm}{\large\bf .\hspace{2mm}
 ASYMPTOTIC PROPERTIES}}
 \end{center}

 \setcounter{equation}{0}

Obviously, asymptotic properties
depend on the
shape of the source term F. Therefore, when $F=0$,
Lemma 3.1 assures that the solution {\em exponentially vanishes} when t
tends to infinity.

When $F=
f(x,t)$ one has:

\vspace{3mm}
{\bf Theorem 4.1}- {\em When the source term } $f(x,t)$ {\em satisfies
condition}:

\beq                   \label{36}
|\, f(x,t)\,| \, \leq\, h \, \frac{1}{(k+t)^{1+\alpha}}\\\  \ \\ \ \ \\ \
\ \ \ \ \ \ \  \ \\ \ \ \  \ \ \
(h, k, \alpha = \hbox{const} > 0),
\eeq

\noindent
{\em then solution } (\ref{34}) {\em  is vanishing
as} t $\rightarrow  \infty$,
{\em at least as} $t^{-\alpha}$.

{\em Moreover, when}

\beq                   \label{37}
|\, f(x,t)\,| \, \leq\, C \, e^{-\delta t},  \ \\ \ \ \\ \\ \ \ \ \ \ \ \ \
\
(C, \delta = \hbox{const} > 0),
\eeq

\noindent
{\em one has}:

\beq                  \label{38}
|\, u(x,t)\,| \, \leq\, k \, e^{-\delta^* t},  \ \\ \ \ \\ \\   \  \ \
\delta^* = min\{\beta, \delta\}, \ \ k=\hbox{const}.
\eeq

{\bf Proof.}
Since
 $u_{g_i }= O(e^{-\beta t})$ (i=0,1),
the proof of the theorem follows directly from the assumptions on
$f(x,t) $.
\hbox{} \hfill \rule{1.85mm}{2.82mm}

\vspace{1mm}
As for the {\em non linear} problem, previous estimates can be applied.
For instance, if we refer to PSGE, it results:

\beq                                    \label {39}
|F(x,t,u)|=|\sin u  +\gamma| \ \leq \gamma_1,  \ \ \ \ \  \gamma_1 =
\hbox{const}.
\eeq

\noindent
Then, the solution of the related non linear strip
problem is {\em bounded for all t}.

Results concerning the exponential decay can be obtained when stronger
assumptions on the decreasing properties of
the source  F(x,t,u(x,t)) are imposed. For instance, according to
[\ref{ce}], when

\beq
|F(x,t,u(x,t))| \leq \hbox{const}. \ \  e^{-\mu t} \ \ \ \ \\\\\\\\\\\\ (\mu
>0),
\eeq

\noindent
then, the solution u(x,t) of the {\em non linear} problem (\ref{12}) vanishes exponentially,
 due to properties (\ref{21}) of G.

\vspace{5mm}
\begin{center}
\large \bf{REFERENCES}
 \end{center}
\vspace{3mm}

\begin{enumerate}
{\footnotesize

\item  H.Lamb, {\it Hydrodynamics}, Dover Publ.inc,
(1932)\label{lam}

\vspace*{-3mm}
\item  D.D. Joseph, M. Renardy, J. Saut, {\it Hyperbolicity and Change
of Type in the Flow of Viscoelastic Fluids}, Arch.Rational Mech.
Analysis, 213-251
(1985).\label{jrs}

\vspace*{-3mm}
\item  A.Morro, L.E.Payne, B Straughan, {\it Decay growth,
countinuous dependence and uniqueness results of generalized heat
conduction theories}, Appl,Anal.38, 231-243 (1990)\label{mps}

\vspace*{-3mm}
\item  A. C. Scott, F. Y. Chu, S. A. Reible, {\it Magnetic-flux
  propagation on a Josephson transmission}, J. Appl. Phys. 47, (7)
3272-3286 (1976).\label{sco}

\vspace*{-3mm}
\item S. Pagano, {\it Licentiate Thesis DCAMM}, Reports 42, Teach
    Univ. Denmark Lyngby Denmark, (1987).
\label{pag1}

\vspace*{-3mm}
\item  M. Tinkham, {\it Introduction to Superconductivity},
     McGraw-Hill,
     (1996).\label{tin}

\vspace*{-3mm}
\item  M.Cyrot D. Pavuna {\it Introduction to Superconductivity
and High-$T_C$ Materials}, World Scientific
(1991).\label{cp}

\vspace*{-3mm}

\item  M. G. Forest, S. Pagano, R. D. Parmentier, P. L. Christiansen,
   M. P. Soerensen, S. P. Sheu, {\it Numerical evidence for global bifurcations
   leading to switching phenomena in long Josephson junctions}, Wave Motion,
    12, 213-226 (1990).\label{for}

\vspace*{-3mm}
\item  P.S.Lomdahl, H.Soerensen, P.L. Christiansen, A.C.Scott, J.C.Eilbeck
 {\it Multiple frequency generation by bunched solitons in
Josephson tunnel junctions }, Phy Rew B 24,12  7460-7462
         (1981).\label{lom1}

\vspace*{-3mm}
\item B. D'Acunto, M. De Angelis, P. Renno, {\it Fundamental
solution of a dissipative operator}, Rend. Acc. Sc. Fis. Mat. 295-314
(1997).\label{ddr}

\vspace*{-3mm}
\item  B. D'Acunto, M. De Angelis, P. Renno, {\it Estimates for
the perturbed Sine Gordon equation
}, Rend. Cir Mat Pal. serieII Supp 57, 199-203 (1998).\label{bari}

\vspace*{-3mm}
\item  B. D'Acunto, P. Renno, {\it On the operator $\varepsilon
\partial_{xxt}+c^2\partial_{xx}-\partial_{tt}$ in general domains
,} Atti Sem Mat Fis Univ Modena ,  XLVII, 191-202 (1999). \label{dr}

\vspace*{-3mm}
\item  T.K. Caughey, J. Ellison, {\it Existence, uniqueness and
stability of solutions of a class of non linear
partial differential
equation}, J.Math Anal. Appl. 51, 1-32 (1975).
\label{ce}

}
$$$$\end{enumerate}

\end{document}